\documentclass[12pt]{article}
\pdfoutput=1

\usepackage{amsmath,amssymb}
\usepackage{graphicx}

\makeatletter

\renewcommand\section{\@startsection{section}{1}{\z@}
                                   {-3.5ex \@plus -1ex \@minus -.2ex}
                                   {2.3ex \@plus .2ex}
                                   {\normalfont\large\bfseries}}
\renewcommand\subsection{\@startsection{subsection}{2}{\z@}
                                   {-3.25ex\@plus -1ex \@minus -.2ex}
                                   {1.5ex \@plus .2ex}
                                   {\normalfont\normalsize\bfseries}}
\renewcommand\subsubsection{\@startsection{subsubsection}{3}{\z@}
                                   {-3.25ex\@plus -1ex \@minus -.2ex}
                                   {1.5ex \@plus .2ex}
                                   {\normalfont\normalsize\bfseries}}
\renewcommand\paragraph{\@startsection{paragraph}{4}{\z@}
                                   {3.25ex \@plus1ex \@minus.2ex}
                                   {-1em}
                                   {\normalfont\normalsize\bfseries}}

\makeatother

\newcommand{\beq}{\begin{equation}}
\newcommand{\eeq}{\end{equation}}
\newcommand{\bea}{\begin{eqnarray}}
\newcommand{\eea}{\end{eqnarray}}

\newcommand{\SL}{{\rm SL}}
\newcommand{\SU}{{\rm SU}}
\newcommand{\SO}{{\rm SO}}

\newcommand{\C}{\mathbb C}

\newcommand{\R}{\mathbb R}

\newcommand{\id}{\hbox{1\kern-.27em l}}

\newcommand{\ad}{{\rm ad}}

\newcommand{\cO}{{\cal O}}

\begin{document}

\pagestyle{empty}

\begin{center}

\vspace*{30mm}
{\Large 't~Hooft Operators in the Boundary}

\vspace*{30mm}
{\large M{\aa}ns Henningson}

\vspace*{5mm}
Department of Fundamental Physics\\
Chalmers University of Technology\\
S-412 96 G\"oteborg, Sweden\\[3mm]
{\tt mans@chalmers.se}     
     
\vspace*{30mm}{\bf Abstract:}
\end{center}
We consider a topologically twisted maximally supersymmetric Yang-Mills theory on a four-manifold of the form $V = W \times \R_+$. 't~Hooft disorder operators localized in the boundary component at finite distance of $V$ are relevant for the study of knot theory on the three-manifold $W$, and have recently been constructed for a gauge group of rank one. We extend this construction to an arbitrary gauge group $G$. For certain values of the magnetic charge of the 't~Hooft operator, the solutions are obtained by embedding the rank one solutions in $G$ and can be given in closed form.

\newpage \pagestyle{plain}

\section{Introduction}
Maximally supersymmetric Yang-Mills theory in four dimensions admits a topological twisting\footnote{This particular twisting is an element of a $\C P^1$ family of inequivalent twistings \cite{Kapustin-Witten,Witten2011}; the generalization has also been used in \cite{Gaiotto-Witten}. There are also two further unrelated possible twistings \cite{Yamron, Vafa-Witten}.} which leads to localization equations of the form
\bea \label{d=4_equations}
F - \phi \wedge \phi + * d_A \phi & = & 0 \cr
d_A (* \phi) & = & 0 
\eea
together with\footnote{This second set of equations (\ref{second}) typically forces $\sigma$ to vanish identically, and will not be considered further in this note.}
\bea \label{second}
d_A \sigma & = & 0 \cr
[ \phi, \sigma] & = & 0 \cr
[ \sigma, \overline{\sigma} ] & = & 0 .
\eea
Here $d_A$ is the covariant exterior derivative associated to a connection $A$ with field strength $F = d A + A \wedge A$ on the gauge bundle $E$ (a principal $G$-bundle over the four-manifold $V$ on which the theory with gauge group $G$ is defined.) The other bosonic fields are a one-form $\phi$ and a complex zero-form $\sigma$ with values in the vector bundle $\ad (E)$ associated to $E$ via the adjoint representation of $G$. There is a Lie product understood in the $\phi \wedge \phi$ term, and $*$ denotes the Hodge duality operator induced from the Riemannian structure on $V$.

As described in \cite{Witten2011}, on an open four-manifold $V$ of the form 
\beq
V = W \times \R_+ ,
\eeq
these equations are relevant to the theory defined on a stack of coincident $D3$-branes terminating on a $D5$-brane. They must then be supplemented by suitable boundary conditions at both ends of $V$. These have been described in \cite{Witten2011} and further elaborated in \cite{Henningson}. With $0 < y < \infty$ a linear coordinate on $\R_+$, the boundary conditions at infinity state that
\beq
A + i \phi \rightarrow \rho
\eeq
as $y \rightarrow \infty$, where $\rho$ is a fixed flat connection on the complexification $E_\C$ of $E$. The boundary conditions at finite distance are related to an embedding of the tangent frame bundle of $W$ as a sub-bundle of $\ad (E)$ via a `principal embedding' of $\SO (3)$ in $G$ \cite{Kostant}. Denoting the corresponding images of the vielbein and the spin connection of $W$ as $e$ and $\omega$ respectively, we have the `Nahm-pole' behavior
\bea \label{Nahm-pole}
A & \rightarrow & \omega \cr
\phi - \frac{1}{y} e & \rightarrow & 0
\eea
as $y \rightarrow 0^+$.

For a generic closed curve $\gamma$ in $V = W \times \R_+$, it is not possible to construct a line operator supported on $\gamma$ and invariant under the topological supersymmetry. But such operators do exist for $\gamma$ of the form
\beq
\gamma =  K \times \{ 0 \} ,
\eeq
where $K$ is a closed curve in $W$.  't~Hooft operators of that kind are relevant for the gauge-theory approach to knot theory developed in \cite{Witten2011} and aimed at making contact with the invariants given by the Jones polynomial \cite{Jones} and Khovanov homology \cite{Khovanov}. These operators are labelled by the highest weight $w$ of a representation of the Langlands dual $G^\vee$ of $G$. On the complement of $K$ in $W$, the solution is equivalent to the solution in the absence of the 't~Hooft operator up to a `large' gauge transformation. The topological class of this gauge transformation is determined by $w$, and for non-trivial $w$ it cannot be extended over $K$. Together with the requirement that the solution be non-singular in the interior of $V$ this determines the asymptotic boundary behaviour completely. 

For the case when $G$ is of rank one, i.e. $G = \SU (2)$ or $G = \SO (3)$, explicit model solutions with these properties were determined in \cite{Witten2011} for arbitrary weights $w$. The purpose of this note is to analyze the case of a general $G$. We hope that this may be useful for performing explicit calculations along the lines of \cite{Gaiotto-Witten}.

In the next section, we will describe an Ansatz that respects the symmetries of the problem, and in section three we will discuss how the required boundary behavior determines a particular solution. We will arrive at a fairly good qualitative understanding, although it is only for certain special weights $w$ that exact solutions (obtained by embedding of the rank one solutions) can be given in closed form. 

\section{The Ansatz}
We take $W = \C \times \R$ so that 
\beq
V = \C \times \R \times \R_+ ,
\eeq
which we endow with the standard metric
\beq
d s^2 = |d z|^2 + d x^2 + d y^2 .
\eeq
(Here $z$, $x$, and $y$ are standard coordinates on the three factors.)  The 't~Hooft operator will be localized along
\beq
K = \{ 0 \} \times \R \times \{ 0 \} , 
\eeq
i.e. at $z = y = 0$.

By a choice of gauge and a certain vanishing theorem \cite{Kapustin-Witten, Witten2011, Witten2010}, the components of $A$ and $\phi$ respectively in the direction of $\R_+$ vanish. Furthermore, we make the Ansatz that the component of $A$ in the direction of $\R$ vanishes and that the solution is invariant under translations along $\R$. The remaining variables are thus
\bea
A & = & A_z d z + A_{\bar{z}} d \bar{z} \cr
\phi & = & \phi_z d z + \phi_{\bar{z}} d \bar{z} + \phi_x d x ,
\eea
and depend on $z$, $\bar{z}$ and $y$ only. In terms of the components of $A$ and $\phi$, the equations (\ref{d=4_equations}) read
\bea \label{holomorphic}
\partial_y A_{\bar{z}} & = & D_{\bar{z}} \phi_x \cr
D_{\bar{z}} \phi_z & = & 0 \cr
\partial_y \phi_z & = & - \left[ \phi_x, \phi_z \right]
\eea
together with
\beq \label{moment_map}
- \partial_y \phi_x = 2 F_{z \bar{z}} + \frac{1}{2} \left[\phi_z, \phi_{\bar{z}} \right] .
\eeq 

We postpone the treatment of the `moment map' equation (\ref{moment_map}) for a while, and start by considering the `holomorphic' equations (\ref{holomorphic}). They can be solved by temporarily interpreting $\phi_x$ as the component of the gauge field in the $y$-direction, and are then invariant under gauge transformations with a parameter valued in the complexification $G_\C$ of $G$. Their content is that the covariant derivatives in the $y$ and $\bar{z}$-directions annihilate $\phi_z$ and commute with each other, so the general solution is
\bea
\phi_z & = & g \varphi g^{-1} \cr
\phi_x & = & - \partial_y g g^{-1} \cr
A_{\bar{z}} & = & - \partial_{\bar{z}} g g^{-1} .
\eea
Here $\varphi = \varphi (z)$ is an arbitrary holomorphic function with values in the the Lie algebra of $G_\C$, and the gauge transformation parameter $g = g (z, \bar{z}, y)$ is an arbitrary function with values in $G_\C$.  

Away from the locus $z = 0$, the Nahm-pole boundary condition (\ref{Nahm-pole}) corresponding to a principal embedding requires $\varphi$ to lie in the `regular nilpotent orbit'. (See e.g \cite{Collingwood-McGovern}). At $z = 0$, $\varphi$ must then lie in the closure of the regular nilpotent orbit, but it may define a more special nilpotent conjugacy class. To describe the possibilities, we choose a Cartan torus $T$ with Lie algebra ${\bf t}$ in $G$ and a principal embedding of $\SO (3)$ in $G$ with standard generators $J^1, J^2, J^3$  such that $J^3 \in {\bf t}$. The commutation relations of $J^+ = J^1 + i J^2$, $J^- = J^1 - i J^2$, and $J^3$ are
\bea
[J^3, J^+] & = & J^+ \cr
[J^3, J^-] & = & - J^- \cr
[J^+, J^-] & = & 2 J^3 .
\eea
We now take
\beq \label{hw}
\varphi = h J^+ h^{-1} ,
\eeq
where
\beq
h \colon \C^* \rightarrow T_\C
\eeq
is a  holomorphic homomorphism such that $\varphi$ has no pole at $z = 0$. (Here $T_\C$ is the complexification of $T$.) This means that
\beq \label{h}
h = \exp (w \log z) ,
\eeq
where $w$ is an element of the weight lattice of the Langlands dual group $G^\vee$ (normalized so that $\exp (2 \pi i w) = 1$) subject to a certain non-negativity condition. In fact, there is a one-to-one correspondence (up to conjugation) between such $w$ and highest weight representations of $G^\vee$. A solution with this $\varphi$ defines what we mean by a 't~Hooft operator in the corresponding representation inserted at $z = 0$ in the boundary $y = 0$. 

As an example, we consider the case where $G = \SU (n)$ so that $G_\C = \SL (n, \C)$. We choose $T$ and $T_\C$ to consist of diagonal unimodular $n \times n$ matrices with complex entries that are of unit modulus or just non-zero respectively. An arbitrary holomorphic homomorphism $h \colon \C^* \rightarrow T_\C$ is then of the form
\beq
h = \left( \begin{matrix}
z^{w_1} & 0 & \ldots & 0 \cr
0 & z^{w_2} & \ldots & 0 \cr
\vdots & \vdots & \ddots & \vdots \cr
0 & 0 & \ldots & z^{w_n}
\end{matrix} \right)
\eeq
with integers $w_1, w_2, \ldots, w_n$ subject to 
\beq
w_1 + w_2 + \ldots + w_n = 0 . 
\eeq
Defining the principal embedding by
\bea
J^3 & = & \frac{1}{2} \left( \begin{matrix}
n - 1 & 0 & \ldots & 0 \cr
0 & n - 3 & \ldots & 0 \cr
\vdots & \vdots & \ddots & \vdots \cr
0 & 0 & \ldots & - (n - 1)
\end{matrix} \right) \cr
J^+ & = & \left( \begin{matrix} 
0 & \sqrt{1 (n - 1)} & 0 & \ldots & 0 \cr 
0 & 0 & \sqrt{2 (n - 2)} & \ldots & 0 \cr
\vdots & \vdots & \vdots & \ddots & \vdots \cr
0 & 0 & 0 & \ldots & \sqrt{(n - 1) 1} \cr
0 & 0 & 0 & \ldots & 0 
\end{matrix} \right) \cr
J^- & = & \left( J^+ \right)^\dagger ,
\eea
we get
\beq
\varphi = \left( \begin{matrix}
0 & \sqrt{1 (n - 1)} z^{w_1 - w_2} & 0 & \ldots & 0 \cr 
0 & 0 &  \sqrt{2 (n - 2)} z^{w_2 - w_3} & \ldots & 0 \cr
\vdots & \vdots & \vdots & \ddots & \vdots \cr
0 & 0 & 0 & \ldots & \sqrt{(n - 1) 1} z^{w_{n - 1} - w_n} \cr
0 & 0 & 0 & \ldots & 0 
\end{matrix} \right) ,
\eeq
so regularity at $z = 0$ amounts to the non-negativity conditions
\beq \label{non-negativity}
w_1 \geq w_2 \geq \ldots \geq w_n . 
\eeq
The number of saturated inequalities  in (\ref{non-negativity}) determines precisely which nilpotent orbit appears at $z = 0$; the trivial case when $w_1 = w_2 = \ldots = w_n = 0$ gives the regular nilpotent orbit, and of course corresponds to a trivial 't~Hooft operator.

We now return to the general case and turn our attention to the remaining moment map equation (\ref{moment_map}).  Together with the boundary conditions, this will determine $g$ uniquely up to an ordinary $G$-valued gauge transformation. By exploiting this gauge symmetry, it is sufficient to consider $g$ of the form 
\beq \label{g}
g = e^{u - (w + J^3) \log | z |} ,
\eeq
where $u = u (z, \bar{z}, y)$ is an element of the Lie algebra ${\bf t}$ of the Cartan torus $T$ of $G$. \footnote{Since there is no factor of $i$ in the exponent, $g$ is not an element of $T$ or even of $G$ but only of $G_\C$.} We then have
\bea
\phi_x & = & - \partial_y u \cr
\phi_z & = & | z |^{-1} e^{u + \frac{w}{2} \log \frac{z}{\bar{z}}} J^+ e^{- u - \frac{w}{2} \log \frac{z}{\bar{z}}} \cr
A_{\bar{z}} & = & - \partial_{\bar{z}} u + \frac{1}{2} (w + J^3) \bar{z}^{-1} ,
\eea
and the moment map equation (\ref{moment_map}) reads\footnote{Note that the right hand side is an element of ${\bf t}$ and in particular commutes with the element $e^{\frac{w}{2} \log \frac{z}{\bar{z}}}$ of $T$.}
\bea
\left(4 \partial_z \partial_{\bar{z}} + \partial_y^2 \right) u & = & | z |^{-2} \frac{1}{2} \left[ e^u J^+ e^{-u}, e^{-u} J^- e^u \right] .
\eea
This equation is invariant under rotations of the $z$-plane around the origin, and also under scaling of $y$ and $z$ by a common real positive factor\footnote{These transformations generate the subgroup of the conformal group of $V$ that leaves the boundary and the locus of the 't~Hooft operator invariant.}. We seek a model solution that is invariant under such transformations, which means that $u$ may only depend on $z$, $\bar{z}$, and $s$ in the combination 
\beq
s = | zÊ| / y . 
\eeq
With this Ansatz, the moment map equation is equivalent to a system of ordinary differential equations:
\beq \label{ODEs}
\left( \left(s \frac{d}{d s} \right)^2 + \left(s^2 \frac{d}{d s} \right)^2 \right) u = \frac{1}{2} \left[ e^u J^+ e^{-u}, e^{-u} J^- e^u \right] .
\eeq
There is clearly a $2 r$-dimensional space of bulk solutions, where $r$  is the rank of $G$. In the next section, we will discuss the relevant solution picked out by the boundary conditions.

\section{The solution}
In the vicinity of the two-dimensional surface in $V$ right above the locus of the 't~Hooft operator, we have $s \rightarrow 0^+$ . In that limit, the general solution to (\ref{ODEs}) behaves as
\beq
u = \alpha \log s + \beta + \cO (s) ,
\eeq
for some parameters $\alpha$ and $\beta$ in ${\bf t}$, that must be chosen such that 
\beq
e^u J^+ e^{-u} = \cO (s) .
\eeq
In fact, regularity of $g$ in this limit requires according to (\ref{g}) that
\beq
\alpha = w + J^3
\eeq
so that
\bea
e^u J^+ e^{-u} & = & s e^{w \log s + \beta} J^+ e^{-w \log s - \beta} \cr
& = & \cO (s)
\eea
by the non-negativity condition on the weight $w$. For a given $w$, the boundary condition as $s \rightarrow 0^+$ thus leaves us with a codimension $r$ space of solutions to (\ref{ODEs}) parametrized by $\beta$.

In the vicinity of the boundary of $V$, we have $s \rightarrow \infty$. In that limit, the Nahm-pole boundary condition requires that
\beq \label{large_s}
u = J^3 \log s + \cO (s^{-1}) .
\eeq
Linearizing (\ref{ODEs}) around such a solution gives the equation
\beq
\left( \left(s \frac{d}{d s} \right)^2 + \left(s^2 \frac{d}{d s} \right)^2 \right) \tilde{u} = s^2 \left(\frac{1}{2} \left[J^-, [J^+, \tilde{u}] \right] +\frac{1}{2} \left[J^+, [J^-, \tilde{u}] \right] + \cO (s^{-1}) \tilde{u} \right)  
\eeq
for the first order deviation $\tilde{u}$. To analyze this equation, we note that
\beq
\frac{1}{2} \left[J^-, [J^+, \tilde{u}] \right] + \frac{1}{2} \left[J^+, [J^-, \tilde{u}] \right] = \left[J^3, [J^3, \tilde{u}] \right] + \frac{1}{2} \left[J^-, [J^+, \tilde{u}] \right] + \frac{1}{2} \left[J^+, [J^-, \tilde{u}] \right]
\eeq
is given by the adjoint action of the $\SO (3)$ quadratic Casimir operator
\beq
C = J^3 J^3 + \frac{1}{2} J^+ J^- + \frac{1}{2} J^- J^+
\eeq
on $\tilde{u}$. The eigenvalues of this action of $C$ are of the form $j (j + 1)$, where the $r$ possible integer values of the spin $j$ are those that appear in the decomposition of the adjoint representation of $G$ under the the principally embedded $\SO (3)$. These possible $j$-values (known as the exponents) are given in table 1 for all simple $G$.
\begin{table}
$$
\begin{array}{lll}
\mathrm{algebra} & \mathrm{dimension} & \mathrm{exponents} \cr
\hline
A_r & r^2 + 2 r & 1, \ldots, r \cr
B_r & 2 r^2 + r & 1, 3, \ldots, 2 r - 1 \cr 
C_r & 2 r^2 + r & 1, 3, \ldots, 2 r - 1 \cr
D_r & 2 r^2 - r & 1, 3, \ldots, 2 r - 3, r -1 \cr
E_6 & 78 & 1, 4, 5, 7, 8, 11 \cr
E_7 & 133 & 1, 5, 7, 9, 11, 13, 17 \cr
E_8 & 248 & 1, 7, 11, 13, 17, 19, 23, 29 \cr
F_4 & 52 & 1, 5, 7, 11 \cr
G_2 & 14 & 1, 5 .
\end{array}
$$
\caption{Dimensions and exponents of simple Lie algebras}
\end{table}
The spin $j$ component $\tilde{u}_j$ of $\tilde{u}$ should thus obey
\beq
\left( \left(s \frac{d}{d s} \right)^2 + \left(s^2 \frac{d}{d s} \right)^2 \right) \tilde{u}_j = s^2 \left( j (j + 1) + \cO (s^{-1}) \right) \tilde{u}_j .
\eeq
Two linearly independent solutions behave as $s^j$ and $s^{-j - 1}$ respectively for large $s$. Only the latter is acceptable in view of (\ref{large_s}), which leaves us with a codimension $r$ space of solutions of (\ref{ODEs}).

Taking the conditions in both limits $s \rightarrow 0^+$ and $s \rightarrow \infty$ into account should generically give a discrete set of solutions to (\ref{ODEs}). Indeed, for a given weight $w$ we expect to find a unique solution. The singular behavior of this scale and rotationally invariant model solution defines the 't~Hooft operator, but further non-singular terms are allowed to appear when the 't~Hooft operator is inserted in a more complicated configuration.

When $w$ is a multiple of $J^3$, i.e. when
\beq
w = k J^3
\eeq
for some non-negative integer $k$, the model solution is given by embedding the rank one solution of \cite{Witten2011} in $G$ and can be given in closed form: We then have
\beq
u = f J^3 ,
\eeq
where the real function $f$ obeys 
\beq
\left( \left(s \frac{d}{d s} \right)^2 + \left(s^2 \frac{d}{d s} \right)^2 \right) f = e^{2 f} .
\eeq
This ordinary differential equation has a two-dimensional space of solutions, but imposing that
\beq
f = (k + 1) \log s + \mathrm{finite}
\eeq
as $s \rightarrow 0^+$ and 
\beq
f = \log s + \cO (s^{-1})
\eeq
as $s \rightarrow \infty$ determines $f$ uniquely:
\beq
f = \log \frac{2 (k + 1) s^{k + 1}}{(\sqrt{1 + s^2} + 1)^{k + 1} - (\sqrt{1 + s^2} - 1)^{k + 1}} .
\eeq
For a more general weight $w$, it appears that the model solution can only be determined numerically.

\vspace*{5mm}
This research was supported by grants from the G\"oran Gustafsson foundation and the Swedish Research Council.

\end{document}